\documentclass[twocolumn]{aastex63}
\usepackage{appendix}
\usepackage{longtable}
\usepackage{amsmath}
\usepackage{chemformula}

\shorttitle{Detection of \ch{(CHOH)2} in the ISM}
\shortauthors{Rivilla et al.}

\begin{document}

\title{Precursors of the RNA-world in space: \\ Detection of ($Z$)-1,2-ethenediol in the interstellar medium, a key intermediate in sugar formation
}

\author[0000-0002-2887-5859]{V\'ictor M. Rivilla}
\affiliation{Centro de Astrobiolog\'ia (CSIC-INTA), Ctra. de Ajalvir Km. 4, Torrej\'on de Ardoz, 28850 Madrid, Spain}
\affiliation{INAF-Osservatorio Astrofisico di Arcetri, Largo Enrico Fermi 5, 50125, Florence, Italy}


\author[0000-0001-8064-6394]{Laura Colzi}
\affiliation{Centro de Astrobiolog\'ia (CSIC-INTA), Ctra. de Ajalvir Km. 4, Torrej\'on de Ardoz, 28850 Madrid, Spain}
\affiliation{INAF-Osservatorio Astrofisico di Arcetri, Largo Enrico Fermi 5, 50125, Florence, Italy}

\author[0000-0003-4493-8714]{Izaskun Jim\'enez-Serra}
\affiliation{Centro de Astrobiolog\'ia (CSIC-INTA), Ctra. de Ajalvir Km. 4, Torrej\'on de Ardoz, 28850 Madrid, Spain}

\author[0000-0003-4561-3508]{Jes\'us Mart\'in-Pintado}
\affiliation{Centro de Astrobiolog\'ia (CSIC-INTA), Ctra. de Ajalvir Km. 4, Torrej\'on de Ardoz, 28850 Madrid, Spain}

\author[0000-0002-6389-7172]{Andr\'es Meg\'ias}
\affiliation{Centro de Astrobiolog\'ia (CSIC-INTA), Ctra. de Ajalvir Km. 4, Torrej\'on de Ardoz, 28850 Madrid, Spain}

\author[0000-0002-6492-5921]{Mattia Melosso}
\affiliation{Department of Chemistry ``Giacomo Ciamician'', University of Bologna, Via F. Selmi 2, Bologna, 40126, Italy}
\affiliation{Scuola Superiore Meridionale, Universit\`{a} di Napoli Federico II, Largo San Marcellino 10, Naples, 80138, Italy}

\author[0000-0002-9953-8593]{Luca Bizzocchi}
\affiliation{Department of Chemistry ``Giacomo Ciamician'', University of Bologna, Via F. Selmi 2, Bologna, 40126, Italy}

\author[0000-0001-6049-9366]{\'Alvaro L\'opez-Gallifa}
\affiliation{Centro de Astrobiolog\'ia (CSIC-INTA), Ctra. de Ajalvir Km. 4, Torrej\'on de Ardoz, 28850 Madrid, Spain}

\author[0000-0001-5191-2075]{Antonio Mart\'inez-Henares}
\affiliation{Centro de Astrobiolog\'ia (CSIC-INTA), Ctra. de Ajalvir Km. 4, Torrej\'on de Ardoz, 28850 Madrid, Spain}

\author[0000-0002-7387-9787]{Sarah Massalkhi}
\affiliation{Centro de Astrobiolog\'ia (CSIC-INTA), Ctra. de Ajalvir Km. 4, Torrej\'on de Ardoz, 28850 Madrid, Spain}

\author[0000-0002-4782-5259]{Bel\'en Tercero}
\affiliation{Observatorio Astron\'omico Nacional (OAN-IGN), Calle Alfonso XII, 3, 28014 Madrid, Spain}

\author[0000-0002-5902-5005]{Pablo de Vicente}
\affiliation{Observatorio de Yebes (OY-IGN), Cerro de la Palera SN, Yebes, Guadalajara, Spain}

\author[0000-0002-2929-057X]{Jean-Claude Guillemin}
\affiliation{Univ Rennes, \'Ecole Nationale Sup\'erieure de Chimie de Rennes, CNRS, ISCR – UMR6226, F-35000 Rennes, France}

\author[0000-0001-6484-9546]{Juan Garc\'ia de la Concepci\'on}
\affiliation{Centro de Astrobiolog\'ia (CSIC-INTA), Ctra. de Ajalvir Km. 4, Torrej\'on de Ardoz, 28850 Madrid, Spain}

\author[0000-0002-5351-3497]{Fernando Rico-Villas}
\affiliation{Centro de Astrobiolog\'ia (CSIC-INTA), Ctra. de Ajalvir Km. 4, Torrej\'on de Ardoz, 28850 Madrid, Spain}

\author[0000-0003-3721-374X]{Shaoshan Zeng}
\affiliation{Star and Planet Formation Laboratory, Cluster for Pioneering Research, RIKEN, 2-1 Hirosawa, Wako, Saitama, 351-0198, Japan}

\author[0000-0001-9281-2919]{Sergio Mart\'in}
\affiliation{European Southern Observatory, Alonso de C\'ordova 3107, Vitacura 763 0355, Santiago, Chile}
\affiliation{Joint ALMA Observatory, Alonso de C\'ordova 3107, Vitacura 763 0355, Santiago, Chile}

\author{Miguel A. Requena-Torres}
\affiliation{University of Maryland, College Park, ND 20742-2421 (USA)}
\affiliation{Department of Physics, Astronomy and Geosciences, Towson University, Towson, MD 21252, USA}

\author[0000-0002-9555-7834]{Francesca Tonolo}
\affiliation{Department of Chemistry "Giacomo Ciamician", University of Bologna, Via F. Selmi 2, Bologna, 40126, Italy}
\affiliation{Scuola Normale Superiore, Piazza del Cavalieri 7, 56126, Pisa, Italy}

\author[0000-0003-3152-3261]{Silvia Alessandrini}
\affiliation{Department of Chemistry "Giacomo Ciamician", University of Bologna, Via F. Selmi 2, Bologna, 40126, Italy}
\affiliation{Scuola Normale Superiore, Piazza del Cavalieri 7, 56126, Pisa, Italy}

\author[0000-0002-1009-7286]{Luca Dore}
\affiliation{Department of Chemistry "Giacomo Ciamician", University of Bologna, Via F. Selmi 2, Bologna, 40126, Italy}

\author[0000-0001-6420-4107]{Vincenzo Barone}
\affiliation{Scuola Normale Superiore, Piazza del Cavalieri 7, 56126, Pisa, Italy}

\author[0000-0002-2395-8532]{Cristina Puzzarini}
\affiliation{Department of Chemistry "Giacomo Ciamician", University of Bologna, Via F. Selmi 2, Bologna, 40126, Italy}




\begin{abstract}

We present the first detection of ($Z$)-1,2-ethenediol, \ch{(CHOH)2}, the enol form of glycolaldehyde, in the interstellar medium towards the G+0.693-0.027 molecular cloud located in the Galactic Center. 
We have derived a column density of (1.8$\pm$0.1)$\times$10$^{13}$\,cm$^{-2}$, which translates into a molecular abundance with respect to molecular hydrogen of 1.3$\times$10$^{-10}$. 
The abundance ratio between glycolaldehyde and ($Z$)-1,2-ethenediol is $\sim$5.2\@.
We discuss several viable formation routes through chemical reactions from precursors such as \ch{HCO}, \ch{H2CO}, \ch{CHOH} or \ch{CH2CHOH}.
We also propose that this species might be an important precursor in the formation of glyceraldehyde (\ch{HOCH2CHOHCHO}) in the interstellar medium through combination with the hydroxymethylene (\ch{CHOH}) radical.

\end{abstract}






\section{Introduction}
\label{sec:intro}

One of the most prominent hypotheses to explain the origin of life is the so called RNA-world (\citealt{gilbert1986}), in which RNA molecules could have performed at the same time the genetic and metabolic functions that nowadays are carried out by DNA and proteins, respectively. 
In the last decade, several prebiotic chemistry experiments in the laboratory have shown that the building blocks of RNA, ribonucleotides, can be formed from simple organic molecules (\citealt{powner2009,patel2015,becker2016,becker2019}). 
Were these simple RNA-world precursors available on early Earth? 
While some authors have discussed the possibility of an endogenous formation on the planet (e.g. \citealt{ruiz-mirazo2014,benner2019}), a possible (and complementary) origin is extraterrestrial delivery (\citealt{oro1961,chyba1992,cooper2001,pasek2008,pearce2017,rubin2019,rivilla2020}) during the heavy cometary and meteoritic bombardment that occurred around 3.9\,Gyr ago (\citealt{marchi2014}).
In this way, a broad repertoire of molecules could have landed on Earth’s surface from outer space, which implies that they may have already been formed in the interstellar medium (ISM) in the natal molecular cloud prior to the formation of the Solar System.

Among the key molecular precursors of the RNA-world, sugars have a prominent role. \citet{powner2009} proposed a viable formation route of RNA nucleotides in prebiotic conditions starting from simpler sugars, like glycolaldehyde (\ch{HOCH2CHO}) and glyceraldehyde (\ch{HOCH2CHOHCHO}). New laboratory experiments by \citet{becker2019} have also shown that glycolaldehyde is a key intermediate in the formation of RNA nucleotides. 
From all this, a prerequisite for the RNA-world is the availability of simple sugars.
Glycolaldehyde has been detected in the ISM in different sources
(\citealt{hollis_interstellar_2000, requena-torres_largest_2008,beltran_first_2009,jorgensen_detection_2012}),
and several works have addressed its formation routes through interstellar chemistry (e.g. \citealt{bennett2007,woods2012,woods2013,butscher2015,fedoseev2015,coutens2018,skouteris2018,rivilla2019a}). 
While glyceraldehyde has been discovered in meteoritic samples (\citealt{cooper2001,pizzarello2012,furukawa2019}), and in laboratory experiments of interstellar ice analogues (\citealt{meinert2016,demarcellus2015,fedoseev2017}), its interstellar detection is elusive despite observational searches (e.g. \citealt{jimenez-serra2020}). 

Recently, the experiments on interstellar ice analogues carried out by \citet{kleimeier2021} have identified the enol form of glycolaldehyde, ($Z$)-1,2-ethenediol \ch{(CHOH)2}, a species that might be a key precursor in the formation of sugars in both interstellar and prebiotic chemistry (e.g. \citealt{kitadai2018}).
However, 1,2-ethenediol has never been searched for in the ISM, since its rotational spectrum was unknown until recently. In this Letter we use the new laboratory spectroscopic measurements performed by \citet{melosso2022diol} to provide the first detection of \ch{(CHOH)2} in the ISM towards the G+0.693$-$0.027 molecular cloud (G+0.693, hereafter) located in the  Sgr B2 complex within the Galactic Center.

\begin{table*}
\centering
\tabcolsep 4.5pt
\caption{List of observed transitions of ($Z$)-1,2-ethenediol \ch{(CHOH)2}. We provide the transitions frequencies, quantum numbers, base 10 logarithm of the integrated intensity at 300 K (log $I$), the values of $S_{\rm ul}\times\mu^2$, the base 10 logarithm of Einstein coefficients (log $A_{\rm ul}$), and upper state degeneracy ($g_{\rm u}$). We also report the derived velocity-integrated intensities of each transition, the rms noise of the spectrum, and the signal-to-noise ratio of the transition (see text).  
The last column gives the information about the species whose transitions are partially blended with the \ch{(CHOH)2} lines.}
\begin{tabular}{ c c c c c c c c c c l}
\hline
 Frequency & Transition  & log $I$ & $S_{\rm ul}\times\mu^2$ & log  $A_{\rm ul}$  & $g_{\rm u}$ & $E_{\rm u}$ & $\int{T_A^* d{\rm v}}^{a}$ & rms &  S/N & Blending$^b$ \\
 (GHz) & ($J_{K_a,K_c},\,v$)  &   (nm$^2$ MHz) & (D$^2$) & (s$^{-1}$) & & (K) & (mK km s$^{-1}$) & (mK) & &  \\
\hline
35.4628492     & 3$_{1,2}$,\,0 - 2$_{1,1}$,\,1 &  -6.0616  & 10.1267 & -6.9024 &  42 &  4.1 &  38.0 & 0.58 & 10.4 & unblended \\
35.4635659     & 3$_{1,2}$,\,1 - 2$_{1,1}$,\,0  & -5.8397  & 10.1267 & -7.1244 &  70 &  4.1 &  63.4 & 0.58 & 17.3 &  unblended \\ 
41.0160144     & 4$_{1,4}$,\,0 - 3$_{1,3}$,\,1  & -5.7895  & 14.2307 & -6.6743 &  54 &  5.6 &  51.3 & 0.70 & 12.6 &  unblended \\ 
41.0167312     & 4$_{1,4}$,\,1 - 3$_{1,3}$,\,0  & -5.5676  & 14.2308 & -6.8962 &  90 &  5.6 &  85.4 & 0.70  & 20.9 &  unblended \\ 
43.1064014     & 4$_{0,4}$,\,0 - 3$_{0,3}$,\,1  & -5.4996  & 15.0483 & -6.8072 &  90 &  5.3 &  100.0 & 0.54 & 32.2 &  blended with U \\ 
43.1071218     & 4$_{0,4}$,\,1 - 3$_{0,3}$,\,0  & -5.7214  & 15.0482 & -6.5854 &  54 &  5.3 &   60.2 & 0.54 & 19.4 &  blended with U \\ 
44.2240556     & 4$_{2,2}$,\,1 - 3$_{2,1}$,\,0  & -5.8240  & 11.3946 & -6.6728 &  54 &  8.0 &   32.6 & 0.70 & 8.2  &  unblended \\
44.2247658     & 4$_{2,3}$,\,1 - 3$_{2,2}$,\,0  & -5.6021  & 11.3946 & -6.8945 &  90 &  8.0 &   54.3 & 0.70 & 13.7 &  unblended \\
47.1196699     & 4$_{1,3}$,\,0 - 3$_{1,2}$,\,1  & -5.4482  & 14.2237 & -6.7158 &  90 &  6.4 & 90.4 & 0.90 & 18.3  &  unblended \\ 
47.1203898     & 4$_{1,3}$,\,1 - 3$_{1,2}$,\,0  & -5.6701  & 14.2237 & -6.4938 &  54 &  6.4 & 54.3 & 0.90 & 11.0 &  unblended \\ 
72.2429524     & 7$_{0,7}$,\,0 - 6$_{0,6}$,\,1  & -5.0451  & 26.1443 & -5.8945 &  90 & 14.3 & 64.8 & 2.18 & 4.2 &  unblended \\ 
72.2436790     & 7$_{0,7}$,\,1 - 6$_{0,6}$,\,0  & -4.8233  & 26.1436 & -6.1164 & 150 & 14.3 & 108.0 & 2.18 &  7.0 &  unblended \\ 
81.6367065     & 8$_{0,8}$,\,0 - 7$_{0,7}$,\,1  & -4.6642  & 29.8923 & -5.9531 & 170 & 18.2 & 89.3 & 2.92 & 4.6 &  blended with $t$-HCOSH \\  
81.6374347     & 8$_{0,8}$,\,1 - 7$_{0,7}$,\,0  & -4.8861  & 29.8909 & -5.7314 & 102 & 18.2 & 53.6 &  2.92 & 2.7 &  blended with $t$-HCOSH \\ 
87.1257785     & 8$_{2,7}$,\,0 - 7$_{2,6}$,\,1  & -4.8570  & 28.3789 & -5.6692 & 102 & 21.7 & 36.3 &  1.63 &  3.4 &  blended with C$_3$H$_6$\\ 
87.1265029     & 8$_{2,7}$,\,1 - 7$_{2,6}$,\,0  & -4.6351  & 28.3789 & -5.8911 & 170 & 21.7 & 60.5 &  1.63  &  5.7 &  blended with C$_3$H$_6$\\  
94.0314593     & 8$_{2,6}$,\,0 - 7$_{2,5}$,\,1  & -4.5668  & 28.5912 & -5.7883 & 170 & 22.6 & 59.3 & 1.23& 7.7 &  blended with U \\ 
94.0321932     & 8$_{2,6}$,\,1 - 7$_{2,5}$,\,0  & -4.7886  & 28.5912 & -5.5666 & 102 & 22.6 & 35.6 &  1.23 & 4.6 &  blended with U \\ 
\hline 
\end{tabular}
\label{tab:transitions}
{\\ (a) The integrated intensity of the lines has been obtained from the best fit from MADCUBA (see text). }
{\\ (b) U refers to blending with emission from an unknown (not identified) species.}
\end{table*}

\section{Observations} 
\label{sec:observations}

We have used a high-sensitivity unbiased spectral survey performed with the Yebes 40m (Guadalajara, Spain) and the IRAM 30m (Granada, Spain) telescopes. The observations, using position switching mode, were centered at $\alpha$(J2000.0)=$\,$17$^{\rm h}$47$^{\rm m}$22$^{\rm s}$, $\delta$(J2000.0)=$\,-$28$^{\circ}$21$^{\prime}$27$^{\prime\prime}$.
The line intensity of the spectra was measured in units of $T_{\mathrm{A}}^{\ast}$ since the molecular emission toward G+0.693 is extended over the beams (\citealt{requena-torres_organic_2006,requena-torres_largest_2008,zeng2018,zeng2020}).
The IRAM 30m observations cover the range from 71.770 to 116.720\,GHz.
The noise of the spectra (in $T_{\rm A}^{*}$) is 1.3 to 2.8\,mK in the 71 to 90\,GHz range, 1.5 to 5.8\,mK in the 90 to 115\,GHz range, and $\sim$10\,mK in the 115 to 116\,GHz range. More detailed information of these IRAM 30m observations is provided in \citet{rivilla2021a,rivilla2021b}.

We have performed a new deeper spectral survey with the Yebes 40m telescope, as part of the 21A014 project (PI: Rivilla). We observed during 29 different sessions between March and June 2021, with a total observing time of $\sim$125\,hours.
The Nanocosmos Q-band (7\,mm) HEMT receiver was used, which provides ultra broad-band observations (18.5\,GHz) in two linear polarizations (\citealt{tercero2021}). The receiver was connected to 16 FFTS providing a channel width of 38\,kHz. 
We observed two different spectral setups centered at 41.40 and 42.30\,GHz, with a total frequency range of 31.07$-$50.42\,GHz. 
We have performed the initial data inspection and reduction using a Python-based\footnote{https://www.python.org} script that uses the CLASS module of the GILDAS package\footnote{https://www.iram.fr/IRAMFR/GILDAS}. For each observing day, spectra averaging was performed, and we fitted baselines using an iterative method that first masks the more visible lines using sigma-clips and then applies rolling medians and rolling averages, interpolating the masked regions with splines. Spectra were then combined, averaged, and exported to MADCUBA{\footnote{Madrid Data Cube Analysis on ImageJ is a software developed at the Center of Astrobiology (CAB) in Madrid; https://cab.inta-csic.es/madcuba/}} (\citealt{martin2019}).
The comparison of the spectra of the two frequency setups was used to identify possible contamination of lines from the image band. 
Telescope pointing and focus were checked every one or two hours through pseudo-continuum observations.
The final spectra were smoothed to 250\,kHz.
The achieved noise at this spectral resolution is $\sim$0.5$-$1.0 mK in $T_{\mathrm{A}}^{\ast}$ scale.
The half power beam width (HPBW) of the telescope was $\sim$35$-$55$^{\prime\prime}$ in the frequency range observed.

\begin{figure*}
\begin{center}
\includegraphics[width=18cm]{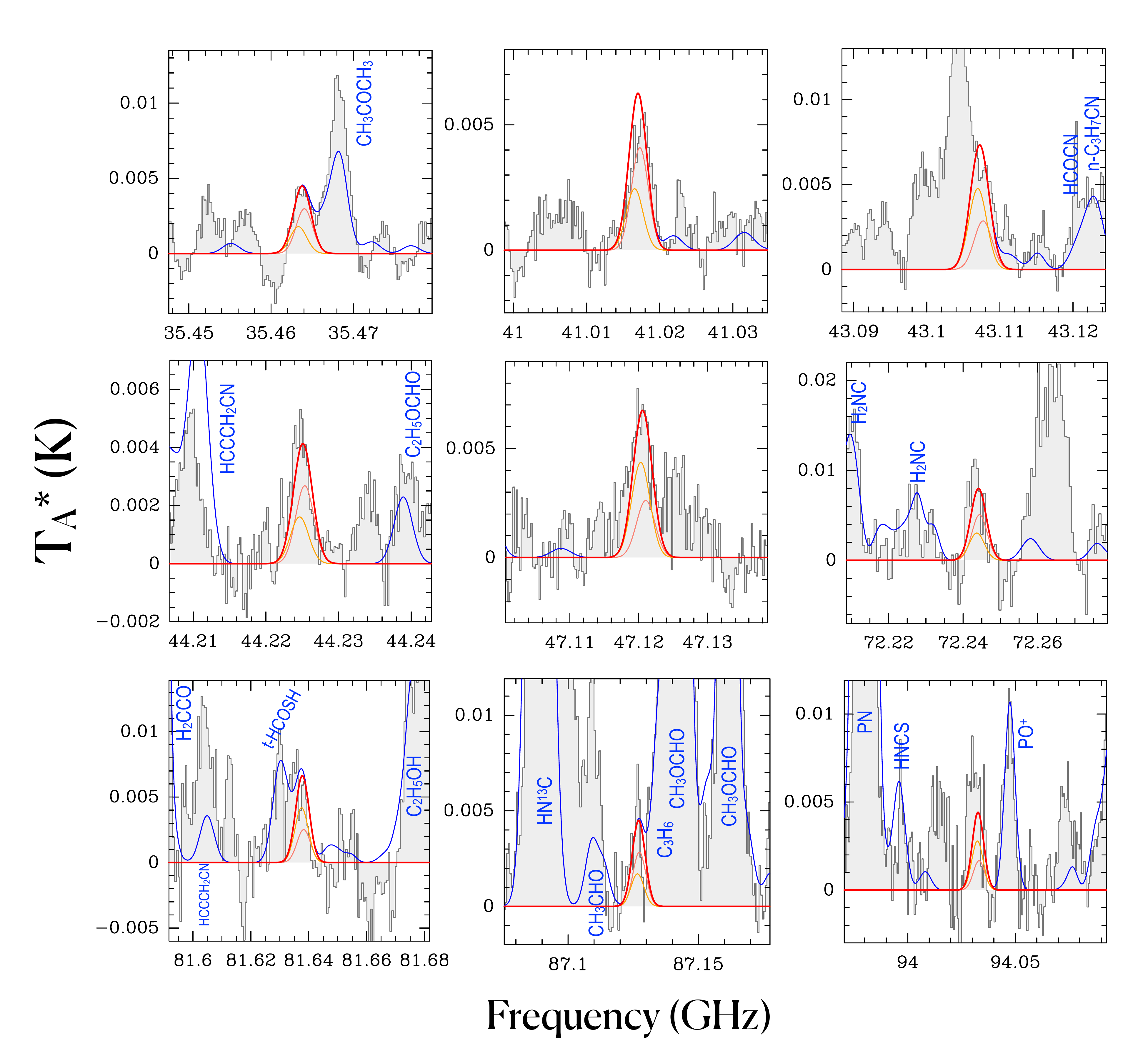}
\vspace{-10mm}
\end{center}
\caption{Selected unblended or slightly blended transitions of 1,2-ethenediol detected towards the G+0.693$-$0.027 molecular cloud. Note that each panel contains two 1,2-ethenediol transitions close in frequency that are not resolved (see Table \ref{tab:transitions}), whose individual contributions are shown with the yellow and salmon curves. The best LTE fit derived with MADCUBA for the 1,2-ethenediol emission is shown with a red curve. The blue curve represents the total emission considering all the species identified towards this molecular cloud.}
\label{fig-spectra}
\end{figure*}

\begin{figure}
\begin{center}
\includegraphics[width=9cm]{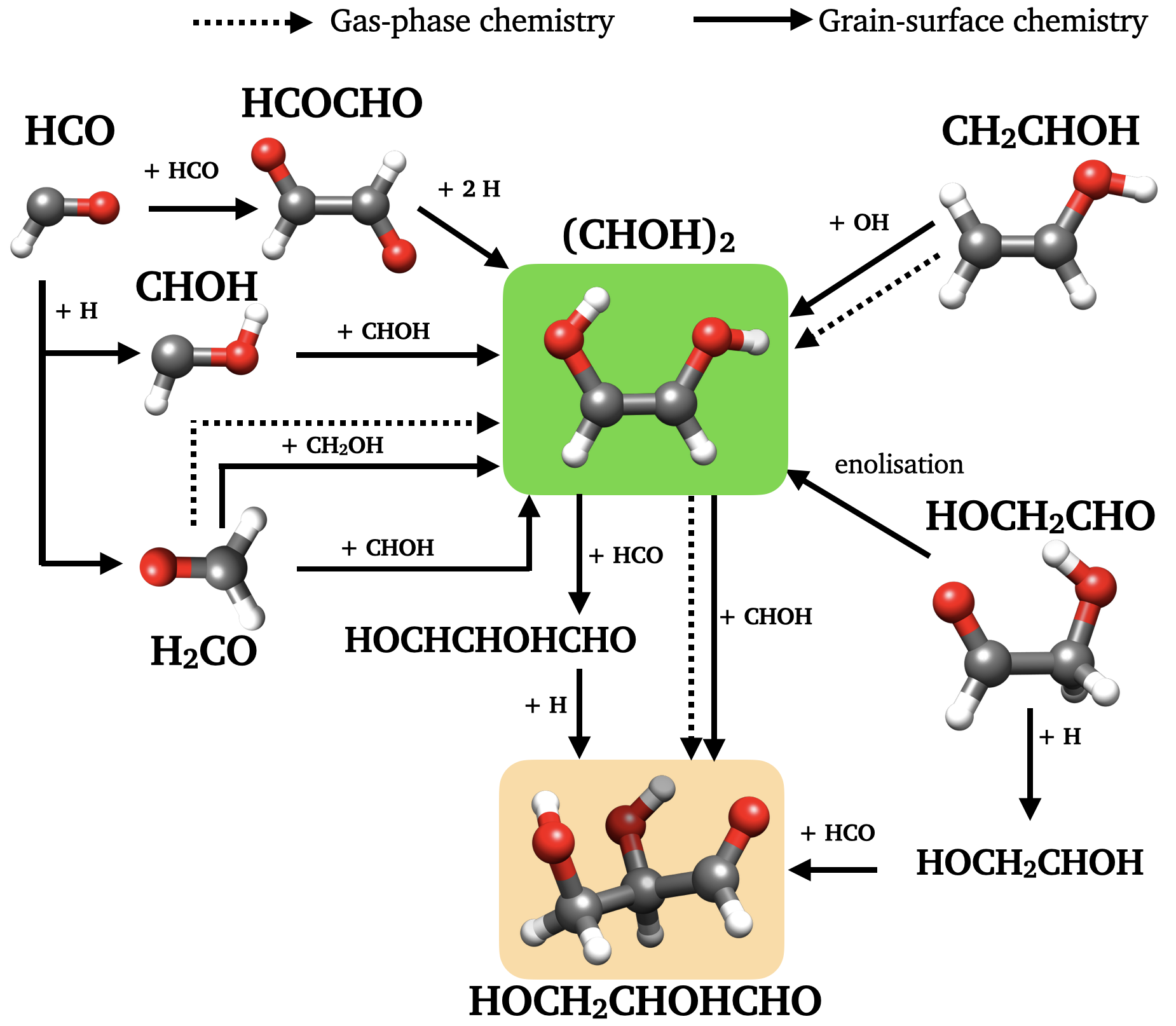}
\end{center}
\vskip-5mm
\caption{Possible formation routes of 1,2-ethenediol (green box), and proposed chemical pathways for the formation of glyceraldehyde (orange box).}
\label{fig:formation}
\end{figure}

\section{Analysis and Results} 
\label{sec:analysis}

To search for transitions of \ch{(CHOH)2} in our spectral survey of the G+0.693 molecular cloud, we have implemented the rotational spectroscopy information from the recent laboratory work by \citet{melosso2022diol} into the SLIM (Spectral Line Identification and Modeling) tool within MADCUBA (version 09/11/2021), which generates synthetic spectra under the assumption of Local Thermodynamic Equilibrium (LTE) conditions. 

Figure~\ref{fig-spectra} presents the eighteen unblended or slightly blended transitions of \ch{(CHOH)2} detected towards G+0.693, arising from nine pairs of unresolved transitions due to the linewidths of this source. Their spectroscopic information is listed in Table~\ref{tab:transitions}.
To properly evaluate possible line contamination by other molecules, we have considered the emission from the $>$ 120 molecules already identified towards G+0.693 (e.g. \citealt{requena-torres_organic_2006,requena-torres_largest_2008,zeng2018,Rivilla2018, rivilla2019b,rivilla2020b,jimenez-serra2020,rivilla2021a,rivilla2021b,rodriguez-almeida2021a,rodriguez-almeida2021b,zeng2021,rivilla2022a}). 
Five pairs of transitions appear unblended, while the other four pairs are slightly blended with thioformic acid ($t$-\ch{HCOSH}, reported in \citealt{rodriguez-almeida2021a}), propene (\ch{C3H6}), and emission from species not yet identified (marked as U in Table \ref{tab:transitions}). 
We note that the rest of the \ch{(CHOH)2} transitions that fall in the spectral coverage of the survey are consistent with the observed spectra, but they are heavily blended with lines from other molecular species.

To derive the physical parameters of the \ch{(CHOH)2} emission, we used the SLIM-AUTOFIT tool of MADCUBA that provides the best non-linear least-squares LTE fit to the data using the Levenberg-Marquardt algorithm. The parameters used in the LTE model are: molecular column density ($N$), excitation temperature ($T_{\rm ex}$), velocity ($v_\text{LSR}$) and full width half maximum (FWHM) of the Gaussian line profiles. 
The fit of \ch{(CHOH)2} profiles was performed by considering also the contribution of the other identified molecules. Since the AUTOFIT algorithm did not converge with all four parameters adjustable, we fixed $v_\text{LSR} =66$\,km\,s$^{-1}$ and FWHM= 20\,km\,s$^{-1}$, which reproduce well the line profiles of the \ch{(CHOH)2} features shown in Figure~\ref{fig-spectra}, and left $N$ and $T_\text{ex}$ as free parameters.
The best LTE fit found by MADCUBA is shown by the red curve in Figure~\ref{fig-spectra}. In each panel, the individual contribution of the two unresolved \ch{(CHOH)2} transitions close in frequency is indicated with the yellow and salmon curves. The blue curve shows the total contribution of all the molecules identified in the survey. 
We obtained a $T_{\rm ex}$=8.5$\pm$0.6\,K, which is in good agreement with the typical values of $T_{\rm ex}$ found in G+0.693, i.e. in the range of 5$-$20\,K (see e.g. \citealt{zeng2018}). The derived column density is (1.8$\pm$0.1)$\times$10$^{13}$\,cm$^{-2}$, which translates into a molecular abundance with respect to molecular hydrogen of 1.3$\times$10$^{-10}$, assuming $N_{\rm H_2}$=1.35$\times$10$^{23}$\,cm$^{-2}$ towards G+0.693 (\citealt{martin_tracing_2008}).

We have estimated the detection level of the transitions of \ch{(CHOH)2} by computing the signal-to-noise ratio derived with the expression:

\vspace{-6mm}
\begin{equation}
\mathrm{S/N} = \mathrm{\Big(\,\int T_A^* dv\Big)\;/\;\Big[rms\,\bigg(\frac{\Delta v}{\mathrm{FWHM}}\bigg)^{0.5}\mathrm{FWHM}\Big]}, \notag
\end{equation}
where $\int{T_A^* d{\rm v}}$ is the velocity-integrated intensity of each line derived from the LTE fit, rms is the noise of the spectrum, and $\Delta$v and FWHM are the spectral resolution and linewidth in velocity units, respectively. The rms was measured around each transition in a nearby line-free spectral range, obtaining values in the 0.5$-$2.5\,mK range. The values of $\Delta$v are 1.5$-$2.5\,km\,s$^{-1}$, while the FWHM used is the one assumed for the LTE fit, i.e. 20\,km\,s$^{-1}$. Table~\ref{tab:transitions} shows that the unblended \ch{(CHOH)2} transitions are detected at S/N levels between 4.6 and 20.9, with at least one of the transitions of the pair $>$ 6.

We have also performed a complementary analysis using the rotational diagram method implemented in MADCUBA (see more details in Appendix \ref{app-0}). Fig. \ref{fig:RD} shows the rotational diagram obtained using the transitions of \ch{(CHOH)2} listed in Table \ref{tab:transitions}. We derived physical parameters fully consistent with the MADCUBA$-$AUTOFIT analysis: $N$=(2.1$\pm$0.5)$\times$10$^{13}$ cm$^{-2}$, and $T_{\rm ex}$=8.6$\pm$1.0 K.

We have also analysed the emission from its isomer glycolaldehyde (\ch{HOCH2CHO}), which is presented in detail in Appendix \ref{app}.
We obtained a column density of (9.3$\pm$0.3)$\times$10$^{13}$ cm$^{-2}$.
The chemically related species (see Section \ref{sec:discussion}) vinyl alcohol (\ch{CH2CHOH}) has also been recently detected towards G+0.693 by Jim\'enez-Serra et al. (submitted). The sum of the column densities of the two conformers identified (\textit{syn} and \textit{anti}) gives 12.4$\times$10$^{13}$ cm$^{-2}$. 
Thus, the ratios between \ch{HOCH2CHO} and \ch{CH2CHOH} with respect to \ch{(CHOH)2} are $\sim$5.2 and $\sim$7.0, respectively.
The upper limit of the molecular abundance of glyceraldehyde reported by \citet{jimenez-serra2020} is $\sim$0.93$\times$10$^{-10}$, only slightly lower than the abundance of \ch{(CHOH)2}, which suggests that deeper spectral surveys are needed to address its detection.

\section{Discussion}
\label{sec:discussion}

\subsection{Formation of 1,2-ethenediol in the ISM}

1,2-ethenediol is the fourth C$_2$H$_4$O$_2$ isomer detected in the ISM, after acetic acid (\ch{CH3COOH}), methyl formate (\ch{CH3OCHO}), and glycolaldehyde (\ch{HOCH2CHO}).  1,2-ethenediol is the highest-energy isomer (\citealt{karton2014pinning}), with the lowest abundance in G+0.693. 
This would be in agreement with the minimum energy principle (MEP, \citealt{lattelais2009}), which states that the most stable isomers (those with lower energy) are the most abundant in the ISM. However, it is well known that the other three isomers do not follow this rule, since the most stable C$_2$H$_4$O$_2$ isomer, acetic acid, is less abundant than methyl formate and glycolaldehyde (see e.g. \citealt{mininni2020} and references therein).
Since the difference in energy of the C$_2$H$_4$O$_2$ isomers is thousands of Kelvin (\citealt{lattelais2009,kua2013}), isomerisation processes seem highly unlikely in the ISM. Therefore, the molecular abundances of the different isomers are expected to be the consequence of kinetics rather than thermodynamics,
namely, they are formed through different chemical pathways that are not directly interconnected.
The chemistry of acetic acid, methyl formate, and glycolaldehyde has been extensively discussed in several works (see e.g. \citealt{laas2011,burke2015,skouteris2018,El-Abd2019,ahmad2020,mininni2020,paulive2021}). 

In this section, we focus our discussion on the newly discovered C$_2$H$_4$O$_2$ isomer,  1,2-ethenediol, whose formation mechanisms have not been investigated so far. A number of viable routes can be hypothesized, which are depicted in Figure~\ref{fig:formation}.
For example, neutral--radical reactions involving formaldehyde and vinyl alcohol:
\begin{align}
 \ch{H2CO + CH2OH &-> (CHOH)2 + H  \:, \\
    CH2CHOH + OH &-> (CHOH)2 + H  \:, }
\end{align}
are likely to be barrierless and involve elimination of a H atom, which removes the excess energy, stabilizing the products. Hence, they can take place both in the gas phase and on grain surfaces. 
In these routes, two of the precursors (\ch{H2CO} and \ch{CH2CHOH}) have been detected towards the G+0.693 molecular cloud at relatively high abundances:  $\geq$3$\times$10$^{-9}$ and 9.2$\times$10$^{-10}$ (\citealt{jimenez-serra2020}, and Jim\'enez-Serra et al. submitted; respectively).
The hydroxymethyl radical (\ch{CH2OH}) has not been detected yet in the ISM, although it is expected to be an important intermediate in the formation of numerous interstellar molecules on the surface of dust grains (\citealt{bermudez2017} and references therein), thus making the surface route (1) a viable formation pathway to \ch{(CHOH)2}. 
The hydroxyl radical (\ch{OH}) is very abundant in the whole Sgr B2 region (\citealt{goico2002}), which makes route (2) a plausible formation pathway of \ch{(CHOH)2}.

Other reactions may occur on grain surfaces only (see Figure \ref{fig:formation}), like the dimerisation of hydroxymethylene:

\begin{equation}
 \ch{2 CHOH -> (CHOH)2}  \:, 
\end{equation}

or the neutral-neutral reaction between  hydroxymethylene and formaldehyde:
\begin{equation}
 \ch{CHOH + H2CO   -> (CHOH)2}  \:, 
\end{equation}

Unfortunately, we cannot search for \ch{CHOH} towards G+0.693, since the required spectroscopic information is not available. 

We have also considered another route based on the dimerisation of the formyl radical (\ch{HCO}) to form glyoxal (\ch{HCOCHO}; \citealt{woods2013}) followed by hydrogenation (Figure \ref{fig:formation}):

\begin{equation}
 \ch{2 HCO -> HCOCHO ->[2H] (CHOH)2}  \:. 
\end{equation}

\ch{HCOCHO} has never been detected in the ISM so far. Its lowest energy conformer ({\textit{trans}}), which is the one expected to be detected in interstellar conditions (the \textit{cis} form is 2237\,K higher in energy), has vanishing dipole moment because of symmetry, so it cannot be detected by rotational emission. 
The route (4) was favoured by the experimental work by \citet{kleimeier2021}, in which \ch{HCOCHO} was identified after electron irradiation of methanol-bearing interstellar ice analogues. However, we note that the experiments by \citet{leroux2020} showed that the hydrogenation of \ch{HCOCHO} ices does not lead to the formation of more complex species (e.g. glycolaldehyde or 1,2-ethenediol; Krim, private comm.), but to lighter molecules such as CO and H$_2$CO. Additional laboratory works, complemented by new quantum mechanical calculations, will help to discern the contribution of the chemical routes proposed (1-5) to the formation of  \ch{(CHOH)2}.   

An alternative formation pathway of \ch{(CHOH)2} could be the enolisation of glycolaldehyde  (Figure \ref{fig:formation}). However, the calculations reported by \citet{kua2013} pointed out that this enolisation requires 39\,kJ\,mol$^{-1}$ (4690\,K), which increases to 88\,kJ\,mol$^{-1}$ (10\,600\,K) in water-assisted isomerisation. 
This indicates that \ch{(CHOH)2} cannot be formed through this process in the conditions of the ISM.

\subsection{Formation of complex sugars in the ISM}
\label{sec:discuss:glycer}

Given its prebiotic relevance, the growth of the chemical complexity of sugars in the ISM is widely debated in astrochemistry. 
In the ISM, both \ch{HOCH2CHO} and its enolic form \ch{(CHOH)2} can be regarded as the closest precursors of glyceraldehyde (see Figure~\ref{fig:formation}). 
The laboratory experiments on H bombardment of interstellar ice analogues by \citet{fedoseev2017} showed that glyceraldehyde formation can take place on grain surfaces via successive neutral--radical reactions:
\begin{align}
 \ch{HOCH2CHO&->[H] HOCH2CHOH} \notag \\
 \ch{&->[HCO] HOCH2CHOHCHO}   \:. \notag  
\end{align}
We here propose an alternative and likely more favourable route on grains, which involves the highly reactive species \ch{(CHOH)2} and \ch{CHOH} (see Figure~\ref{fig:formation}): 
\begin{equation}
 \ch{(CHOH)2 + CHOH -> HOCH2CHOHCHO} \:. \notag  
\end{equation}
The advantage of this pathway is that both reactants are less stable than their corresponding keto forms \citep{schreiner2008capture,karton2014pinning}, which favours to overcome the reaction barrier and enhances the overall process exothermicity.

Therefore, the detection of the 2-carbon sugar glycolaldehyde and its enol form in the ISM opens the possibility of detecting more complex sugars. Several interstellar searches of 3-carbons sugars, already identified in meteorites (\citealt{cooper2001}), including the aldose glyceraldehyde\footnote{An aldose is a sugar consisting of a carbon backbone and a carbonyl group at the end of the carbon chain, resulting in an aldehyde group.} (\citealt{hollis2004propenal}), and the ketose\footnote{A ketose is a sugar consisting of a carbon backbone and a carbonyl group within the backbone.} dihydroxyacetone (\citealt{widicus2005,apponi2006}), indicated that they are at least a factor of 5 less abundant than glycolaldehyde. We also searched for these 3-carbon sugars towards G+0.693 (\citealt{jimenez-serra2020}), and obtained that glyceraldehyde and dihydroxyacetone are less abundant than glycolaldehyde by factors $>$7 and $>$13, respectively. This indicates that 1-carbon increase in chemical complexity of sugars implies a molecular abundance drop of at least one order of magnitude, as previously observed in other chemical families like alcohols, thiols, isocyanates (\citealt{rodriguez-almeida2021a,rodriguez-almeida2021b}; Jim\'enez-Serra et al., submitted), and aldehydes (\citealt{sanz-novo2022}).

\subsection{Implications for the RNA-world}

The backbone of RNA is composed by molecules of ribose (c-\ch{C5H10O5}) linked by phosphate groups.
Initially, it was proposed that ribose, along with other sugars, might be formed through the {\it formose} reaction (\citealt{butlerow1861}), starting from aqueous formaldehyde (\ch{H2CO}) and glycolaldehyde (\ch{HOCH2CHO}). However, stability and stereochemical purity problems (\citealt{larralde1995,toxvaerd2013}) cast doubts about the efficiency of formation of sugars in a prebiotic context. Moreover, the formation of RNA nucleotides adding ribose to nucleobases is highly inefficient (\citealt{fuller1972,orgel2004}). As a consequence, alternative synthesis routes are needed. Several prebiotic experiments (\citealt{powner2009,becker2019}) have found viable synthesis of RNA nucleotides starting from simple precursors such as glycolaldehyde and glyceraldehyde. 
In this context, the new detection of ($Z$)-1,2-ethenediol in the ISM reported here is especially relevant for both extraterrestrial and terrestrial environments. First, this species can be a direct precursor of glyceraldehyde in interstellar ices, as discussed in Section \ref{sec:discuss:glycer}. Secondly, it has been proposed that \ch{(CHOH)2} is the key intermediate in the formation of glyceraldehyde starting from glycolaldehyde under plausible early Earth conditions (\citealt{kitadai2018}):
\begin{equation}
 \ch{HOCH2CHO ->}\ch{(CHOH)2 ->} \ch{HOCH2CHOHCHO} \notag
\end{equation}

Therefore, the detection of 1,2-ethenediol (\ch{(CHOH)2}) towards the G+0.693 molecular cloud, along with those of its tautomer glycolaldehyde (\ch{HOCH2CHO}) in several other sources, indicates that immediate molecular precursors of more complex species like glyceraldehyde are available in the ISM.
These compounds, if delivered to the surface of planets and moons through meteoritic and cometary impacts, could have triggered prebiotic chemistry by providing key precursors of RNA nucleotides.

\bibliography{diol}{}
\bibliographystyle{aasjournal}

\section*{ACKNOWLEDGEMENTS} 

We acknowledge the two anonymous reviewers for their careful reading of the manuscript and their useful comments.
We are grateful to the IRAM 30m and Yebes 40m telescopes staff for their help during the different observing runs.
The 40m radio telescope at Yebes Observatory is operated by the Spanish Geographic Institute (IGN, Ministerio de Transportes, Movilidad y Agenda Urbana).
IRAM is supported by INSU/CNRS (France), MPG (Germany) and IGN (Spain). 
V.M.R., L.C. and A.L-G. have received funding from the Comunidad de Madrid through the Atracci\'on de Talento Investigador (Doctores con experiencia) Grant (COOL: Cosmic Origins Of Life; 2019-T1/TIC-15379). 
I.J.-S., J.M.-P., L.C, and A.M. have received partial support from the Spanish project numbers PID2019-105552RB-C41 and MDM-2017-0737 (Unidad de Excelencia Mar\'ia de Maeztu$-$Centro de Astrobiolog\'ia, INTA-CSIC).
A.M. has received support from the Spanish project number MDM-2017-0737-19-2 and grant PRE2019-091471, funded by MCIN/AEI/10.13039/501100011033 and by “ESF Investing in your future".
P.dV. and B.T. thank the support from the European Research Council through Synergy Grant ERC-2013-SyG, G.A. 610256 (NANOCOSMOS) and from the Spanish Ministerio de Ciencia e Innovación (MICIU) through project PID2019-107115GB-C21. B.T. also thanks the Spanish MICIU for funding support from grant PID2019-106235GB-I00.
J.-C.G. thanks the Centre National d'Etudes Spatiales (CNES) for a grant.

\clearpage

\restartappendixnumbering      
\appendix


\section{Rotational diagram of ($Z$)-1,2-ethenediol, \ch{(CHOH)2}}
\label{app-0}

\begin{figure}
\centering
\includegraphics[width=9cm]{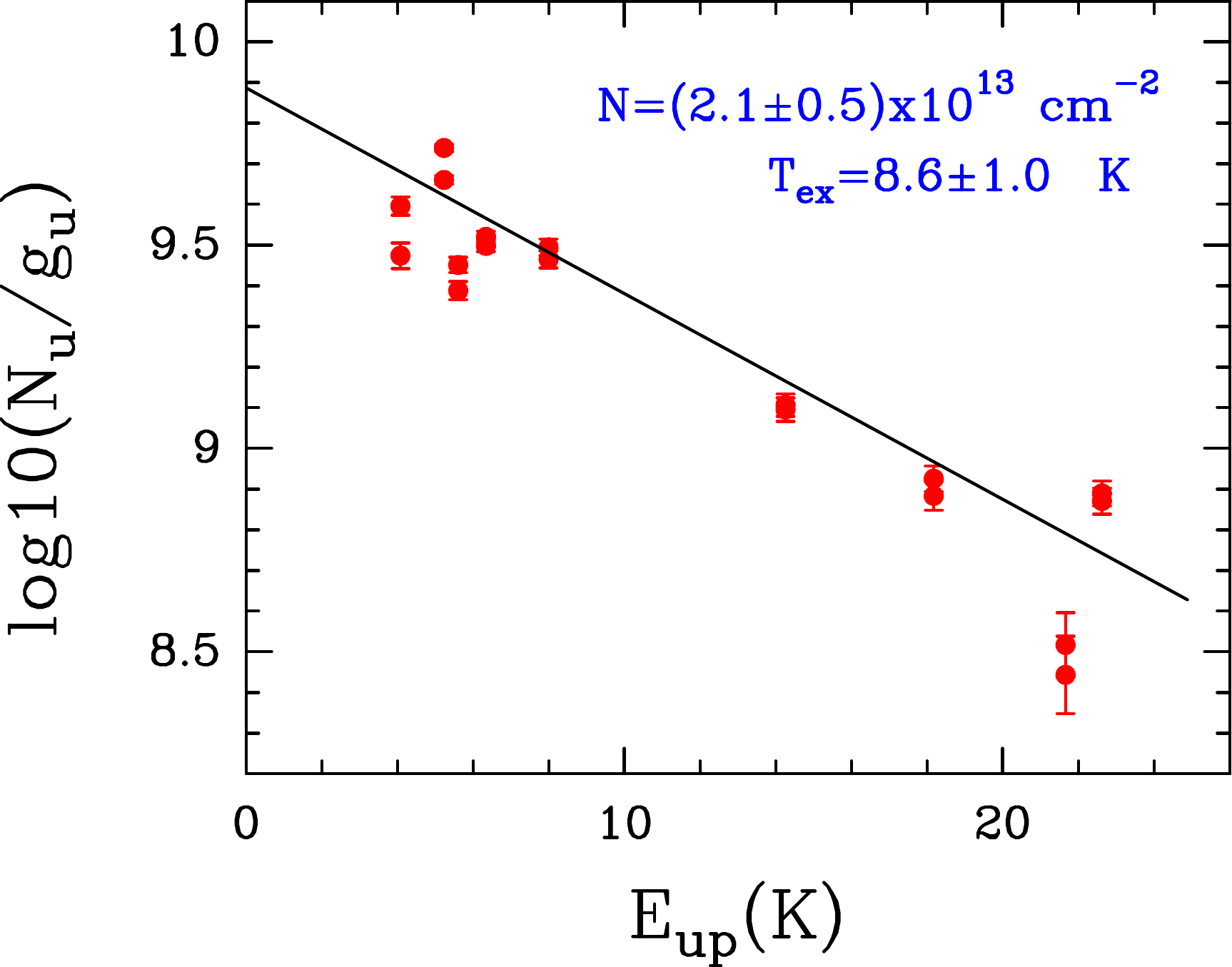}
\caption{Rotational diagram of ($Z$)-1,2-ethenediol. The red dots correspond to the transitions listed in Table  \ref{tab:transitions}. The black line is the best linear fit to the data points. The derived values for the molecular column density ($N$) and the excitation temperature ($T_{\rm ex}$), along with their uncertainties, are indicated in blue in the upper right corner.}
\label{fig:RD}
\end{figure}

We have performed a rotational diagram following the standard procedure (\citealt{goldsmith1999}) implemented in MADCUBA, and described in detail in \citet{rivilla2021a}.
We have used FWHM=20 km s$^{-1}$ and v$_{\rm LSR}$=66 km s$^{-1}$, as in the AUTOFIT analysis.  The resulting diagram is shown in Figure \ref{fig:RD}.
We have obtained $N$=(2.1$\pm$0.5)$\times$10$^{13}$ cm$^{-2}$ and $T_{\rm ex}$=8.6$\pm$1.0 K, which are in agreement within the uncertainties with those derived with AUTOFIT (Section \ref{sec:analysis}).

\section{LTE fit of glycolaldehyde (\ch{HOCH2CHO})}
\label{app}

\begin{figure*}
\begin{center}
\includegraphics[width=40pc]{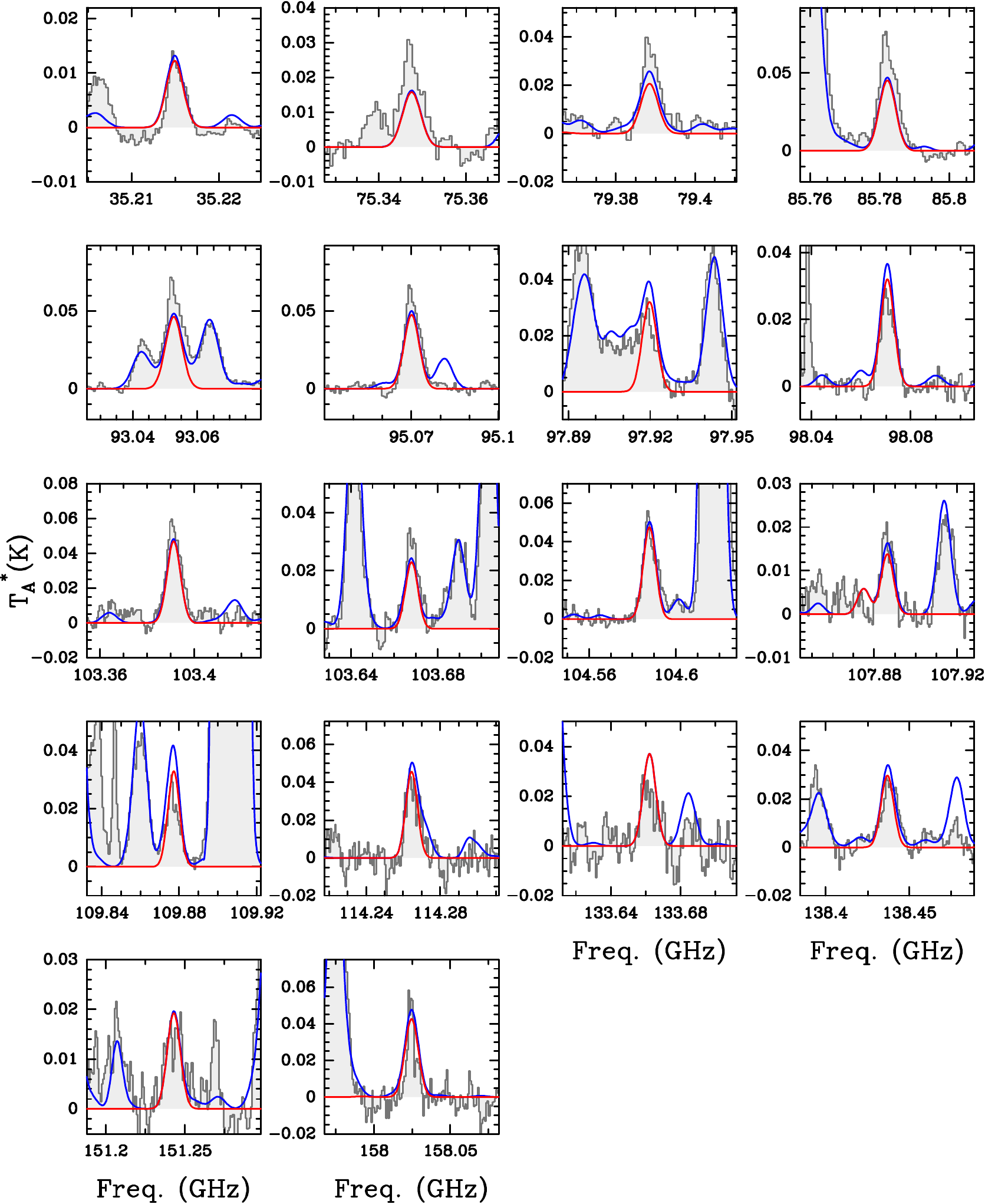}
\end{center}
\caption{Selected unblended transitions of \ch{HOCH2CHO} identified towards the G+0.693$-$0.027 molecular cloud and employed to perform the LTE fit. The corresponding spectroscopic information is provided in Table \ref{tab:glyco}. The best LTE fit derived with MADCUBA is shown with a red curve, while the blue curve depicts the total emission considering all the species identified towards this molecular cloud.}
\label{fig:glyco}
\end{figure*}

Multiple transitions of glycolaldehyde (\ch{HOCH2CHO}) are detected within the spectral survey towards G+0.693. 
We have used the Cologne Database for Molecular Spectroscopy (CDMS, \citealt{endres2016}) entry 060501 (June 2021). 
To perform the LTE fit, we have selected completely unblended transitions that cover a broad range of $E_{\rm up}$ from 4 to 51\,K, so that the $T_{\rm ex}$ can be directly constrained. To evaluate that the transitions are not contaminated, we have used, as explained in the text (Section \ref{sec:analysis}), the line emissions from the all the  molecules identified in the source. We have run MADCUBA using the transitions shown in Table \ref{tab:glyco}. We have left free to vary the four model parameters. 
We obtain $N$=(9.3$\pm$0.3)$\times$10$^{13}$ cm$^{-2}$, $T_{\rm ex}$=21.8$\pm$0.8 K, $v_\text{LSR}$=68.9$\pm$0.2 km s$^{-1}$, FWHM=19.4$\pm$0.5  km s$^{-1}$.
The resulting best LTE fit is shown in red in Figure~\ref{fig:glyco}. 

\begin{table*}[h!]
\centering
\tabcolsep 4.5pt
\caption{Selected transitions of \ch{HOCH2CHO} used to perform the LTE fit. We provide the transitions frequencies, quantum numbers, base 10 logarithm of the integrated intensity at 300\,K (log $I$), the values of $S_{\rm ul}\times\mu^2$, the base 10 logarithm of Einstein coefficients (log $A_{\rm ul}$), and upper state degeneracy ($g_{\rm u}$).}
\begin{tabular}{c c c c c c c }
\hline
 Frequency & Transition  & log $I$ & $S_{\rm ul}\times\mu^2$ & log  $A_{\rm ul}$  & $g_{\rm u}$ & $E_{\rm u}$  \\
 (GHz) & ($J_{K_a,K_c}$)  &   (nm$^2$ MHz) & (D$^2$) & (s$^{-1}$) & & (K)  \\
\hline
 35.2148100  &  $ 6_{1, 5} - 6_{0, 6}$  & -5.4016  &   17.948 &  -6.1538 &  13  &  12.9 \\
 75.3474860  &  $ 8_{1, 7} - 7_{2, 6}$  & -4.8156  &   15.498 &  -5.3430 &  17  &  21.4 \\
 79.3884359  &  $ 4_{2, 3} - 3_{1, 2}$  & -4.9263  &   10.349 &  -5.1741 &   9  &   7.9 \\
 85.7822074  &  $ 8_{1, 8} - 7_{0, 7}$  & -4.3514  &   34.509 &  -4.8265 &  17  &  18.9 \\
 93.0526650  &  $ 9_{0, 9} - 8_{1, 8}$  & -4.2245  &   39.850 &  -4.7062 &  19  &  23.4 \\
 95.0700722  &  $ 9_{1, 9} - 8_{0, 8}$  & -4.2039  &   40.025 &  -4.6763 &  19  &  23.4 \\
 97.9196993  &  $ 3_{3, 1} - 2_{2, 0}$  & -4.6336  &   13.362 &  -4.6807 &   7  &   8.8 \\
 98.0705529  &  $ 3_{3, 0} - 2_{2, 1}$  & -4.6328  &   13.343 &  -4.6792 &   7  &   8.8 \\
103.3913370  &  $10_{0,10} - 9_{1, 9}$  & -4.0826  &   45.462 &  -4.5552 &  21  &  28.4 \\
103.6679627  &  $10_{1, 9} - 9_{2, 8}$  & -4.3248  &   26.204 &  -4.7910 &  21  &  32.0 \\
104.5877362  &  $10_{1,10} - 9_{0, 9}$  & -4.0718  &   45.551 &  -4.5394 &  21  &  28.3 \\
107.8863063  &  $ 5_{2, 3} - 4_{1, 4}$  & -4.9176  &   5.759  &  -5.1162 &  11  &  10.9 \\
109.8771836  &  $ 4_{3, 1} - 3_{2, 2}$  & -4.5319  &   13.495 &  -4.6354 &   9  &  11.0 \\
114.2645516  &  $11_{1,11} -10_{0,10}$  & -3.9527  &   51.066 &  -4.4140 &  23  &  33.8 \\
133.6622379  &  $13_{0,13} -12_{1,12}$  & -3.7493  &   62.041 &  -4.1947 &  27  &  46.2 \\
138.4367820  &  $ 7_{3, 5} - 6_{2, 4}$  & -4.2955  &   15.114 &  -4.5070 &  15  &  21.0 \\
151.2430164  &  $14_{1,13} -13_{2,12}$  & -3.7535  &   49.939 &  -4.1590 &  29  &  58.6 \\
158.0246068  &  $ 6_{4, 2} - 5_{3, 3}$  & -4.0767  &   19.193 &  -4.1687 &  13  &  21.2 \\
\hline 
\end{tabular}
\label{tab:glyco}
\end{table*}

\end{document}